\documentclass[12pt]{iopart}
\usepackage{iopams}
\usepackage{amssymb}
\usepackage{bm}
\usepackage{amssymb}
\usepackage{graphicx,epsfig}

\newcommand{\nn}{\nonumber \\}

\def\da     {\downarrow}
\def\ua     {\uparrow}
\def\rg     {\rangle}
\def\lg     {\langle}

\begin{document}

\title{Coherent operation of coupled superconducting flux qubits }

\author{Mun Dae Kim}

\address{Institute of Physics and Applied Physics, Yonsei University, Seoul 120-749, Korea}
\address{Korea Institute for Advanced Study, Seoul 130-722, Korea}
\ead{\mailto{mdkim@yonsei.ac.kr}}


\begin{abstract}
We study the quantum operation of coupled superconducting flux
qubits under a microwave irradiation. The flux qubits can be
described as magnetic dipole moments in the limit of weak
microwave field amplitude consistent with usual experimental
situations. With the Hamiltonian for coupled qubits under a
microwave field,  we show that  a strong coupling enables to
realize the high performance  controlled-NOT gate operation.  For
practical quantum computing we analyze the effect of microwave on
switching function of  phase-coupled qubits.
\end{abstract}

\maketitle

\section{Introduction}

Superconducting Josephson junction qubits are one of the most
promising candidates for implementing quantum computation because
macroscopic coherence of superconductor is robust against noises
from environment.  Recent experiments
for superconducting charge \cite{Wallraff} and flux \cite{Bertet,Kaku}
qubits have reported much longer coherence times.
%
For  practical quantum computing also the high performance coupled-qubit operations need to
be achieved.
For flux qubits an XY-type of coupling between two qubits has been achieved, where
only a SWAP gate operation has been demonstrated \cite{Niskanen}.
The controlled-NOT (CNOT) gate, as the basic
element of the universal gate \cite{Barenco}, provides the simplest
design for scalable quantum computing.  By using an Ising-type of
coupling between two flux qubits, a CNOT gate operation has been
experimentally demonstrated for inductively coupled flux qubits \cite{Plantenberg} with rather weak coupling.

In this study
we obtain the  Hamiltonian for coupled three-Josephson-junction qubits under a
microwave field in weak microwave amplitude limit,
consistent with usual experimental situations.
The Hamiltonian is written in terms of the
magnetic dipole moment of qubit which constitutes the qubit-microwave coupling constant
in experiment \cite{MARIANTONI} and phenomenological Hamiltonians \cite{Saito}.
For a proper parameter regime for the CNOT gate operation the Hamiltonian is
reduced to a block-diagonalized form.
Each diagonal part of the Hamiltonian corresponds to different
control qubit state. The discriminating oscillations of occupation probability of coupled qubits
give rise to the CNOT gate operation.
With the obtained  Hamiltonian for coupled qubits we show that
strongly coupled qubits  can  achieve  the high performance  CNOT gate operation.
The fluctuation effects of  the microwave field as well as the static magnetic field are discussed.
We consider, for example, phase-coupled flux qubits to achieve a strong coupling.
For  practical quantum computing the phase-coupling scheme of  flux qubits provides a switching function.
We discuss the effects of microwave on the switching operation of phase-coupled qubits.


\section{ Hamiltonian of coupled flux qubits interacting with a microwave field}

When only a static flux is penetrating a flux qubit, the fluxoid
quantization condition for the qubit loop is given by $2\pi n+2\pi
f-\phi_{1}-\phi_{2}-\phi_{3}= 0$ with integer $n$.
The reduced flux is denoted as $f= \Phi_{{\rm st}}/\Phi_0$
with the static external flux $\Phi_{{\rm st}}$
threading the  qubit loop and the superconducting unit
flux quantum $\Phi_0=h/2e$.
Using the relation
$\phi_{1}=2\pi (n+f)-\phi_{2}-\phi_{3}$,
the energy levels of the  qubit is written as
$E_{s}(f)    =    E_{J1}[1 - \cos(2\pi (n+f) - \phi_{2,s} - \phi_{3,s})]
   + E_{J2}(1-\cos\phi_{2,s}) +  E_{J3}(1-\cos\phi_{3,s}),$
where $\phi_{2(3),s}$ with $s\in \{\da,\ua\}$
is the value of $\phi_{2(3)}$ for the state $|s\rangle$ at the local
 minima of $U_{{\rm eff}}$ and depends on $f$.
Here $|\downarrow\rangle$ ($|\uparrow\rangle$) is the diamagnetic (paramagnetic)
current state of the flux qubit.

For the Rabi oscillation of a  flux qubit, a microwave
field with frequency $\omega$ is applied;
$f_\omega (t) =(\Phi_{\rm mw}/\Phi_0)\cos\omega t,$
where $\Phi_{\rm mw}=BS$ with the microwave magnetic field $B$ and
the area of the qubit loop $S$.
Using the fluxoid quantization condition
\begin{eqnarray}
\label{fqc}
2\pi n+2\pi (f+f_\omega (t))-\phi_{1}-\phi_{2}-\phi_{3}= 0,
\end{eqnarray}
the effective potential of the qubit is written as $ U_{{\rm
eff}}(\hat{\bm{\phi}})= E_{J1}[1-\cos(2\pi (f+f_\omega (t)
)-{\phi}_{2}-{\phi}_{3})] +E_{J2}(1-\cos{\phi}_{2})+
E_{J3}(1-\cos{\phi}_{3})$ with $n=0$. Normally, in usual
experiments, the applied microwave fields are in the range of
$\Phi_{\rm mw}\ll \Phi_0$.
Then, the effective potential is written as
\begin{eqnarray}
U_{{\rm eff}}(\hat{\bm{\phi}}) &\approx & E_{J1}[1-\cos(2\pi f-\phi_{2}-\phi_{3})] \\
   &&+   E_{J2}(1-\cos\phi_{2})+ E_{J3}(1-\cos\phi_{3})\nonumber \\
   &&+   2\pi E_{J1}(\Phi_{\rm mw}/\Phi_0) \cos\omega
t\sin(2\pi f-\phi_{2}-\phi_{3}), \nonumber
\end{eqnarray}
where $\phi_{2(3),s}$ can be assumed to be constant in time due to $\Phi_{\rm mw}\ll \Phi_0$.

In the basis of the qubit current sates $\{\left|
\uparrow\right\rangle, \left|\downarrow\right\rangle\}$ one
can obtain the qubit Hamiltonian. The
diagonal components $E_{\omega,s}$ of the qubit Hamiltonian consist
of the static $E_{s}$ and oscillating parts;
$E_{\omega,s}(t,f)\simeq E_{s}(f)-M_{s} B \cos\omega t$
with
\begin{eqnarray}
M_{s}= -SI_c\sin(2\pi f-\phi_{2,s}-\phi_{3,s})=SI,
\end{eqnarray}
where the qubit loop has the circulating current
$I=-I_c\sin\phi_{1}$ with $I_c=2\pi E_{J1}/\Phi_0$ and the
magnetic dipole moment of the qubit is $M_{s}$.
Hence, in terms of the magnetic moment of the qubit loop and
the interaction with the microwave field, the qubit Hamiltonian
is rewritten as
\begin{eqnarray}
{\cal H} = {\cal H}_0-{\bf M}\cdot {\bf B}(t),
 \label{OneH}
\end{eqnarray}
where
${\cal H}_0  = E_{\downarrow}\left|\downarrow\right\rangle\left\langle\downarrow\right| +
E_{\uparrow}\left|\uparrow\right\rangle\left\langle\uparrow\right|
 -  t_q(\left|\downarrow\right\rangle\left\langle\uparrow\right|
+ \left|\uparrow\right\rangle\left\langle\downarrow\right|)$
and $t_q$ is the tunneling amplitude between the two states in the qubit, which
comes from the charging energy of the Josephson junctions \cite{Orlando}.
The interaction between the magnetic dipole moment and the
microwave field is described by
\begin{eqnarray}
 {\bf M}\cdot {\bf B}(t)  =   MB \cos\omega t
(\left|\downarrow\right\rangle\left\langle\downarrow\right| -
\left|\uparrow\right\rangle\left\langle\uparrow\right|),
\end{eqnarray}
where we set  $M_{\da}\approx -M_{\ua} \equiv M$.
Here note that, although the microwave field just threads the qubit loop,
not applied on the qubit directly,
the dipole magnetic moment of qubit and the microwave interact with each other through
the fluxoid quantization condition of Eq. (\ref{fqc}).

For  two coupled flux qubits  we consider
that the left qubit is the control qubit.  In this case the flux $f_L$
is adjusted far away from the degeneracy point so that the tunneling
process $t_{L}$ in the left qubit is negligible, i.e.,
$t_{L}/|E_{\downarrow s}-E_{\uparrow s}|\approx 0$. As a
consequence, the two-qubit Hamiltonian becomes block-diagonalized. Hence
the problem is reduced to that of two independent qubits under a
microwave irradiation,
\begin{eqnarray} \label{TwoH}
{\cal H} \!\! &=&\!\!\!\!  \sum_{s,s'}[E_{ss'}(f_L,f_R)-(M_{Ls}+M_{Rs'}) B \cos\omega t ]
|s,s'\rangle\langle s,s'| \nonumber\\
&&- t_{R}|s,s'\rangle\langle s,-s'|,
\end{eqnarray}
where $E_{ss'}(f_L,f_R)$ is the energy level of coupled qubits,
$M_{L(R)s}$ is the magnetic dipole moment of the left (right) qubit
in $|s\rangle$ state, and
$-s$ denotes the opposite pseudo-spin state of $s \in \{\da,\ua\}$.
Here we set $M_{R\da}\approx -M_{R\ua}\equiv M_R$ and $M_{L\da}\approx -M_{L\ua}\equiv M_L$.

To clearly describe the Rabi-type oscillations, we employ
rotated coordinates for the coupled-qubit states as
$|00\rg =\cos(\theta_\da/2)|\da\da\rg +\sin (\theta_\da/2)|\da\ua\rg,~
|01\rg =-\sin(\theta_\da/2)|\da\da\rg +\cos(\theta_\da/2)|\da\ua\rg,~
|10\rg =\cos(\theta_\ua/2)|\ua\da\rg +\sin (\theta_\ua/2)|\ua\ua\rg,$ and
$|11\rg =-\sin(\theta_\ua/2)|\ua\da\rg +\cos(\theta_\ua/2)|\ua\ua\rg$
with
$\tan\theta_s =2t_{R}/|E_{s\downarrow}-E_{s\uparrow}|.$
%
Let us consider, for example, the case  that the control qubit states is $|\downarrow\rangle$.
Then we see the relations
 \begin{eqnarray}
 \label{relation2}
\left|\da\da\right\rg\left\lg \da\da\right| - \left|\da\ua\right\rg\left\lg \da\ua\right| &   =   &
 \cos\theta_\da(|00\rg\lg 00| - |01\rg\lg 01|) \\ && \hspace*{0.05cm}
-\sin\theta_\da(|01\rg\lg 00| + |00\rg\lg 01|),\nn
\left|\da\da\right\rg\left\lg \da\da\right| + \left|\da\ua\right\rg\left\lg \da\ua\right| &   =   &
|00\rg\lg 00| + |01\rg\lg 01|.
 \end{eqnarray}
If $2t_{R}/|E_{s\downarrow}-E_{s\uparrow}|= 0$, i.e.,
$\sin\theta_\da= 0$,  
the off-diagonal terms in Eq. (\ref{relation2}) does not appear so that the
transition between the qubit states does not occur even for a
resonant microwave field. Hence the tunneling $t_R$
between the states $|\da\da\rangle$
and $|\da\ua\rangle$  plays a key role in responding to the microwave
field.
For the control qubit state $|\uparrow\rangle$ we also perform
the transformation and   obtain the two-qubit Hamiltonian,
\begin{eqnarray}
{\cal H}   &=&   \sum_{{\rho}=0,1}
\Big[{\cal E}_{{\rho}0}(t)|{\rho}0\rangle\langle {\rho}0| +{\cal E}_{{\rho}1}(t)|\rho
1\rangle\langle {\rho}1|\nn &+&    \alpha_\rho M_{R}B \cos\omega t (|\rho
0\rangle\langle \rho 1|+|\rho 1\rangle\langle \rho 0|)\Big],
\label{TwoT}
\end{eqnarray}
where
\begin{eqnarray}
\label{calE}
{\cal E}_{{\rho\rho'}}(t) \!  =\!  {\cal E}^0_{{\rho\rho'}}(f_L,f_R)
-[(-1)^\rho M_{L}+(-1)^{\rho'} \beta_\rho M_{R}] B \cos\omega t \nonumber\\
\end{eqnarray}
with $\alpha_0=\sin\theta_\da, ~\alpha_1=\sin\theta_\ua, ~\beta_0=\cos\theta_\da,$ and $\beta_1=\cos\theta_\ua$.

The Hamiltonian in Eq. (\ref{TwoT}) is valid in
the weak microwave amplitude limit $\Phi_{\rm mw}\ll \Phi_0$.
However, to perform a gate operation via a Rabi oscillation in
experiments,  $\Phi_{\rm mw}$ should satisfy a more strict
condition that the Rabi frequency $\Omega$ is much smaller than the
energy gap $\omega_0$, i.e.,
\begin{eqnarray}
\label{rabi}
\Omega= MB/\hbar \ll \omega_0.
\end{eqnarray}
In this regime, the rotating wave approximation
(RWA) can be  applied and a well-behaved Rabi-type oscillation
can be observed. In other words, the applied microwave field should be in the range of
$\Phi_{\rm mw}=BS \ll \hbar\omega_0/I$.
From the experimental parameters
 for the flux qubits in Ref. \cite{Mooij}, this condition reads
$\Phi_{\rm mw} \ll 10^{-3}\Phi_0$. For the Rabi frequency
$\Omega/2\pi \approx$ 600 MHz which coincides with usual
experimental situations, we find that the corresponding amplitude
$\Phi_{\rm mw} \sim 10^{-4}\Phi_0$
provide a well-behaved Rabi-type oscillation in this study.

\section{Controlled-NOT gate operation}

\begin{figure}[b]
\vspace{5.5cm}
\includegraphics{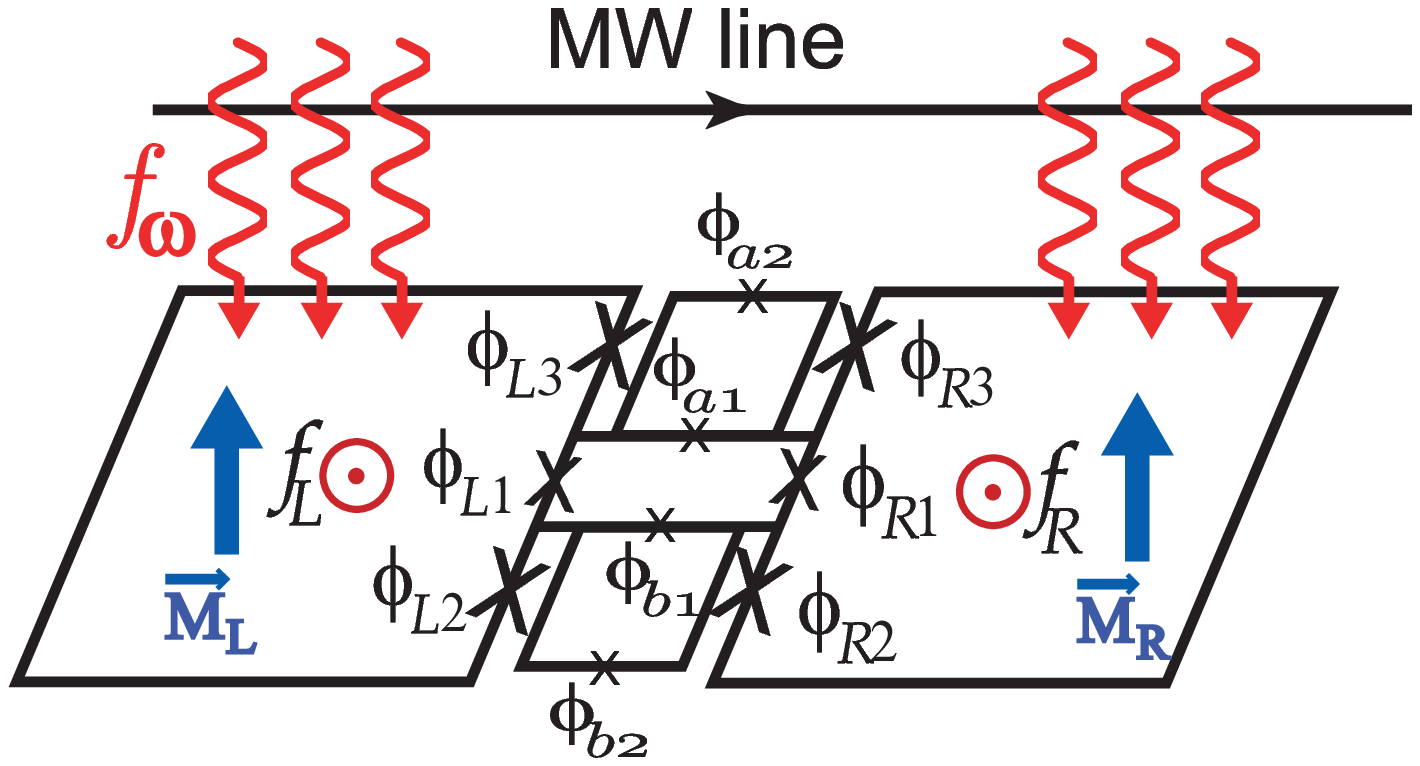}
\vspace{-1cm}
\caption{
Phase-coupled two flux qubits connected by a  loop which has two
dc-SQUIDs providing a switchable coupling between flux qubits. The
Josephson coupling energy of a junction in the dc-SQUIDs with
phase difference $\phi_{a(b)i}$ is $E'_{J}$. The dc-SQUIDs have
threading flux $f_{a(b)}$ which can be adjusted to control the
coupling between two flux qubits. When $f_{a}=f_{b}=0$, a dc-SQUID
can be simplified as a single Josephson junction with the
Josephson coupling energy $2E'_J$. } \label{PC}
\end{figure}

\begin{figure}[b]
%
\vspace{4.5cm}
\includegraphics{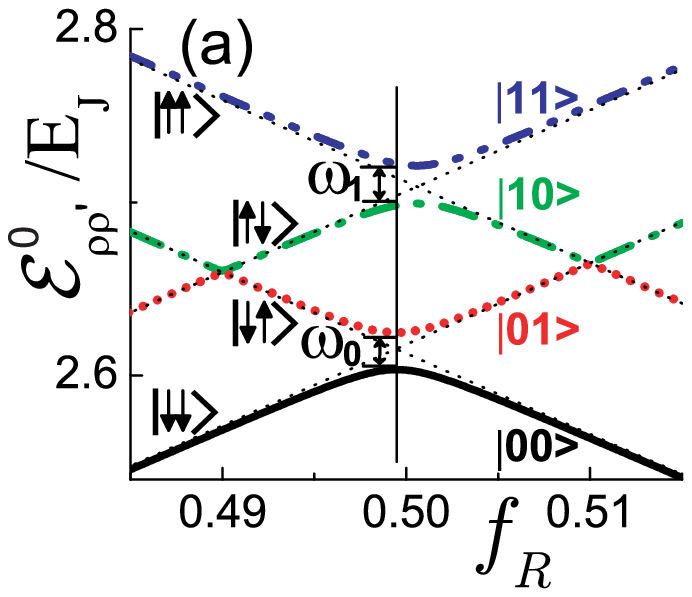}
\includegraphics{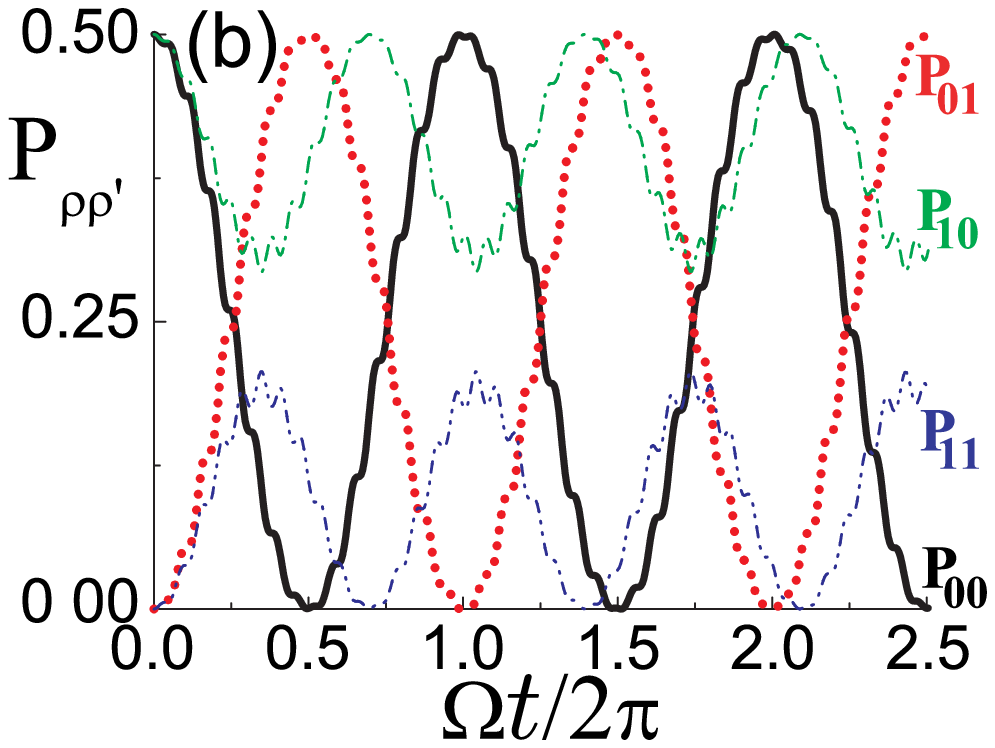}
\vspace{-0.5cm}
%
%
\vspace{4.3cm}
\includegraphics{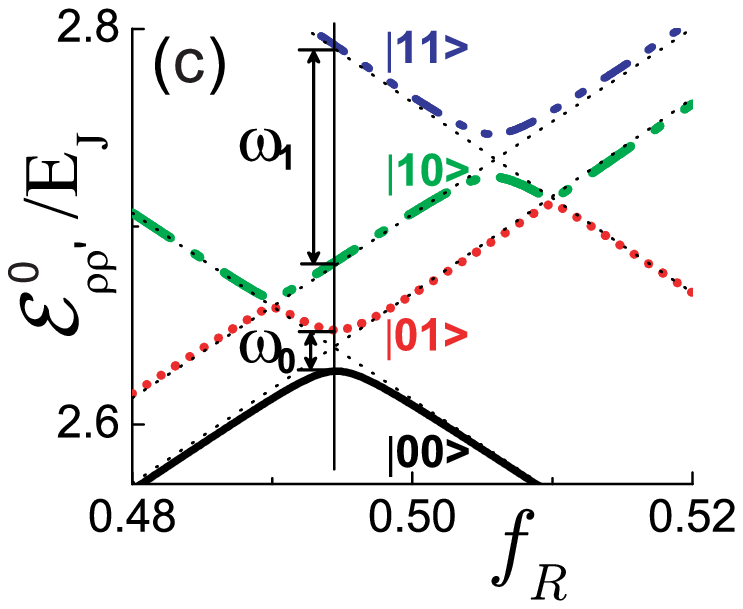}
\includegraphics{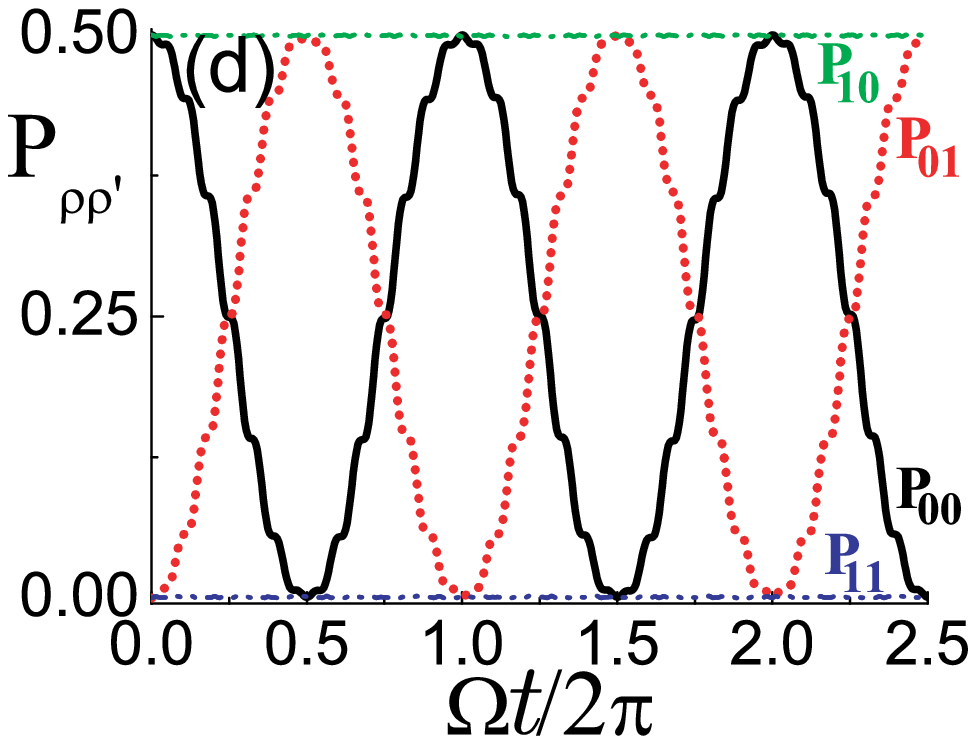}
\vspace{-0.cm}
\caption{ Energy levels ${\cal E}^0_{\rho\rho'}$ of  coupled
qubits in Fig. \ref{PC} for (a) a weak coupling $J/h=$0.6GHz and
(c) a strong coupling $J/h=5$GHz and
occupation probabilities of coupled
flux qubits  during Rabi-type oscillations
at the degeneracy point (b) $f_R \approx 0.4994$
and (d) $f_R \approx 0.4945$ for the weak and strong coupling case, respectively.
Here we choose the parameters as
$f_L=0.49$, $E_{J1}/E_J=0.75$,  and $t_{R}/h$=2GHz.
The initial state  is $\psi(0)=(|00\rangle+|10\rangle)/\sqrt{2}$ and
the Rabi frequency is $\Omega/2\pi=600$MHz.
$E_{ss'}$ with $s,s' \in \{\da,\ua\}$ are shown as thin dotted lines in (a) and (c).
At $\Omega t=$ (odd)$\pi$ the CNOT gate operation is expected to be achieved.
}
\label{level}
\end{figure}

In this section, we consider a concrete system, for example,
two phase-coupled qubits \cite{KimHong,KimTunable,Grajcar3,Ploeg,Grajcar} in Fig. \ref{PC}.
From the Hamiltonian of  Eq. (\ref{TwoT})
we show that the CNOT gate  operation can be achieved  with a high performance for a strong coupling.
In a previous study ~\cite{KimHong}  the CNOT gate operation was analyzed
without  microwave irradiation.
The energy levels ${\cal E}^0_{\rho\rho'}(f_L,f_R)$ of coupled qubits
are shown in Figs. \ref{level}(a) and (c)
as a function of $f_R$ with fixed $f_L=0.49$.

At degeneracy points  (solid lines)  in Figs. \ref{level}(a) and (c),
$E_{\downarrow\downarrow}=E_{\downarrow\uparrow}\equiv E_0$,
$E_{\uparrow\downarrow} < E_{\uparrow\uparrow}$,
$\theta_\downarrow = \pi/2$ and
$\theta_\uparrow = \tan^{-1}(2t_R/|E_{\uparrow\uparrow}-E_{\uparrow\downarrow}|)$.
At this point we have the diagonal elements,
${\cal E}_{00}= -t_{R}-M_{L} B \cos\omega t +E_{0}$,
${\cal E}_{01}=t_{R}-M_{L} B \cos\omega t +E_{0}$,
and the constant energy gap
\begin{eqnarray}
\omega_0&=&{\cal E}_{01}(t)-{\cal E}_{00}(t)=2t_{R}. \label{omega0}
\end{eqnarray}
On the other hand, the off-diagonal term $\alpha_0 M_{R} B \cos\omega t ~(\rho=0)$
with $\alpha_0=1$ gives rise to a dynamical evolution between the states $|00\rangle$ and
$|01\rangle$, i.e., a Rabi-type oscillation.
In Figs. \ref{level} (b) and (d) the occupation
probabilities $P_{\rho\rho'}$ of $|\rho\rho'\rangle$ states during
the Rabi-type oscillations are shown as a function of time when the initial
state is prepared as $\psi(0)=(|00\rangle+|10\rangle)/\sqrt{2}$.

As shown in Fig. \ref{level}, the microwave field with the resonance
frequency $\omega=\omega_0$ induces the Rabi oscillation between the
 states $|00\rangle$ and $|01\rangle$ owing to the off-diagonal term
with $\rho=0$ in Eq. (\ref{TwoT}). For the weak coupling case of  Fig. \ref{level}(b)
the states $|10\rangle$ and $|11\rangle$ also oscillate simultaneously in  response to
the microwave field through the off-diagonal term $\alpha_1 M_{R}B \cos\omega
t (|10\rangle\langle 11|+|11\rangle\langle 10|)$ with $\rho=1$,
while they are stationary for the strong coupling case in Fig. \ref{level}(d).

For a discriminating Rabi oscillation, the coupling strength $J$ should be
larger than the tunneling rate, $J>t_R$.
The coupling strength is given as
$J = (E_{\uparrow\uparrow}-E_{\uparrow\downarrow})/4$
at the degeneracy point where $E_{\downarrow\downarrow}=E_{\downarrow\uparrow}$ \cite{KimHong,KimTunable}.
Further, as discussed in Eq. (\ref{rabi}), for Rabi-type oscillations the
energy gap $\omega_0=2t_R$ should be much larger than the Rabi
frequency $\Omega=M_RB/\hbar$.
Consequently, for a high performance of CNOT gate operation, we see the  criteria
 \begin{equation}
 \Omega \ll \omega_0 (=2t_R) < J.
 \label{criteria}
 \end{equation}
In Fig. \ref{level}(a), at the degeneracy point $f_R \approx
0.4994$, the tunneling amplitude is greater than the coupling
strength, i.e., $t_R/h \approx 2\mathrm{GHz}$ and $J/h\approx 0.6 \mathrm{GHz}$.
Thus, the oscillations are not discriminative. To improve
the discrimination of oscillations,
one need to increase $J$ larger than the value of $t_R$.

From the energy levels in Fig. \ref{level}(c),
 increasing $J$ is shown to make the distance  farther between
 the degeneracy point at $f_R\approx 0.4945$ and the other degeneracy
 point.
 This property makes it possible to perform a discriminative
 Rabi-type oscillation.
 In Fig. \ref{level}(d), we plot the oscillations of
 occupation probabilities $P_{\rho\rho'}$ as a function of time by using
 the Hamiltonian in Eq. (\ref{TwoT}).
 Figure \ref{level}(d) shows that
 the microwave field at the resonance frequency
 $\omega=\omega_0$ induces the Rabi oscillation between the
 states $|00\rangle$ and $|01\rangle$ driven by the off-diagonal term
 for $\rho=0$ in Eq. (\ref{TwoT}), while
 the states $|10\rangle$ and $|11\rangle$ $(\rho=1)$
 do not respond to  the microwave field.
From Eqs. (\ref{calE}) and (\ref{omega0})
 we have the  relation
\begin{eqnarray}
\label{omega1}
\omega_1\!=\!{\cal E}_{11}(t)\!-\!{\cal E}_{10}(t) \!=\!
\sqrt{(4J)^2\!\!+\!\omega^2_0} \!+\! 2\beta_1 M_{R} B \cos\omega t,
\end{eqnarray}
which shows that the difference $|\omega_1-\omega_0|$
 between the energy gaps
 increases as the coupling strength $J$ increases at the degeneracy point.
 Here the the latter time-dependent term is negligibly small compared with $J$
 due to the criteria of Eq. (\ref{criteria}).
Hence the states $|10\rangle$ and $|11\rangle$ hardly respond to the microwave with
frequency $\omega_0$, resulting in the CNOT gate operation
via discriminative Rabi-type oscillations at $\Omega t$=(odd)$\pi$ in Fig. \ref{level}(d).

Now we discuss the effect of  fluctuations of microwave field as well as
the static one on the CNOT gate operation.
During the two-qubit operation
both the static flux $f_{L(R)}$ and microwave flux $f_\omega (t)$
give rise to the noises, $\delta f_{L(R)}(t)$ and
$\delta f_\omega (t)$, which destroy the qubit coherence.
%
These noises appear in the fluxoid quantization condition,
\begin{eqnarray}
\label{bcfluc}
2\pi n+2\pi (f_{L(R)}+f_\omega (t)+\delta f_{L(R)}(t)+\delta f_\omega (t))\nonumber\\
-\phi_{L(R)1}-\phi_{L(R)2}-\phi_{L(R)3}= 0. ~~~~~~~~~~~~~~~~~~
\end{eqnarray}
Let us first discuss the noises in the left qubit.
Combining the fluctuation of static flux $\delta f_{L}(t)$
into that of the microwave field $\delta f_\omega (t)$,
the net effect of both noises is expressed as
a random fluctuation $\delta B_{L}(t)$
in the microwave field $f_\omega(t)=(BS/\Phi_0)\cos\omega t$.
As a result, the magnetic field of microwave in the diagonal terms of Eq. (\ref{TwoT})
is rewritten as $M_{L} (B \cos\omega t+\delta B_{L}(t))$.
Since the diagonal term $M_{L} (B \cos\omega t+\delta B_{L}(t))$
does not change the energy gaps $\omega_0$ and $\omega_1$
in Eqs. (\ref{omega0}) and (\ref{omega1}),
the fluctuation $\delta B_{L}(t)$ has no effect on the operations.
Hence the fluctuations of both the static and microwave fields threading the left qubit loop
are negligible.

On the other hand, the fluctuations of fluxes threading the right qubit
may cause the decoherence in the qubit states.
Let us discuss the terms with $\rho=0$ and $\rho=1$ in Eq. (\ref{TwoT}) separately.
For the  $\rho=0$ terms in  Eq. (\ref{TwoT})
the fluctuation of microwave flux, $\delta f_\omega (t)$,
can be combined into the fluctuation of static flux, $\delta f_R(t)$,
in the fluxoid quantization condition for right qubit in  Eq. (\ref{bcfluc}).
Since the degeneracy point is  optimally
biased with respect to the static flux $f_R$,
the first order fluctuation effect of both  $\delta f_{R}(t)$ and
$\delta f_\omega (t)$ on qubit state dephasing will vanish at this point.

For the terms with $\rho=1$ the net fluctuation  can be given by
$M_{R} (B \cos\omega t+\delta B_{R}(t))$ in the diagonal and off-diagonal terms.
Since $\theta_\uparrow =\tan^{-1}(t_R/2|J|)\approx 0$ and thus $\alpha_1\approx 0$
for a sufficiently strong coupling,
the off-diagonal term with $\rho=1$  will not appear.
Thus the fluctuation $\delta B_{R}(t)$ in  the off-diagonal terms
hardly gives rise to dissipation of qubit states for a sufficiently strong coupling, but
the fluctuations in the diagonal terms may cause dephasing.

\section{ Effect of a microwave on the switching function of coupled qubits}

For  practical quantum computing, the quantum operations should be manipulated by
a switchable coupling. To discuss this, in the model of Fig. \ref{PC}, we
introduce two dc-SQUIDs in the connecting loop of the phase coupled qubits.
Here the external fluxes $f_a$ and $f_b$ vary from zero to 0.5.
When the fluxes threading the dc-SQUID
loops are set as $f_{a}=f_{b}=0$ (switch on), a dc-SQUID can be
simplified as a single Josephson junction with the Josephson
coupling energy $2E'_J$.
The design of Fig. \ref{PC} is similar to that in a previous
 study \cite{KimTunable}. The difference is just the direction of
 pseudo-spin of the right qubit. The present design is symmetric so
 that it may be more appropriate for real experiments.

The current  flowing in the connecting loop $I'$ also gives rise to
magnetic moments in the dc-SQUID loops. $I'$ depends on the two-qubit states such as
$I'=\mp(2\pi/\Phi_0) 2E'_J\cos\pi f'\sin\pi f'_{\rm eff} \equiv \pm
I'_0$ for the states, $|\da\da\rangle$ and $|\ua\ua\rangle$,
respectively and otherwise $I'=0$ \cite{KimTunable}.
Here we set $f_a=f_b=f'.$ Then the Josephson energy of the
connecting loop $U'_{{\rm JJ},\omega}(\phi_{ai},\phi_{bi})
= \sum^2_{i=1}[E'_{J}(1-\cos\phi_{ai}) +
E'_{J}(1-\cos\phi_{bi})]$ has an additional oscillating term,
\begin{equation}
2E'_J 2\pi (BS'/\Phi_0) \cos\pi f'  \sin\pi f'_{\rm eff} \cos\omega t,
\end{equation}
where we used the fluxoid quantization conditions for the dc-SQUID loops and
the connecting loop, $S'$ is the area of the dc-SQUIDs,
and $f'_{\rm eff}\equiv (\phi_{L1}+\phi_{R1})/2\pi$.

As a consequence, the connecting loop energy of coupled states becomes
\begin{eqnarray}
E'_{ss',\omega}=E'_{ss'} -M'_{ss'} B\cos\omega t,
\end{eqnarray}
where $M'_{\da\da}=-M', M'_{\ua\ua}=M'$ and
$M'_{\da\ua}=M'_{\ua\da}=0$. $M'\equiv I'_0S'$ can be interpreted
as the magnetic moment of the control loops. This magnetic moment
arises by the interaction between the magnetic flux and the flowing
current via the fluxoid quantization conditions of the dc-SQUID loops. If we
include these terms in the Hamiltonian of Eq. (\ref{TwoH}), the
net effect is just the shift of magnetic moments,
\begin{equation}
\label{Mc}
M_{sL} \rightarrow M_{sL}+0.5M',~ M_{sR}~ \rightarrow M_{sR}+0.5M',
\end{equation}
remaining the physics qualitatively the same.

The microwave field threading the dc-SQUID loops also generates noises.
The noise  from the microwave field can be introduced as a fluctuation of magnetic field
$0.5M'(B\cos\omega t+\delta B'(t))$ as before. $\delta B'(t)$
can be combined with the previous fluctuations $\delta B_{L(R)}(t)$
through the relation of Eq. (\ref{Mc}),
which does not generate qualitatively different effect.
The static flux $f'$ controls the coupling between two qubits;
when $f'=0 ~(0.5)$, the coupling is switched on (off).
The static flux $f'$ also generates noises in the dc-SQUID loops,
but the switch-on (off) point, $f'=0 ~(0.5)$,  is  an optimal point with respect to $f'$ \cite{KimTunable}.

\section{Discussions and  summary}

\begin{figure}[t]
\vspace{5cm}
\includegraphics{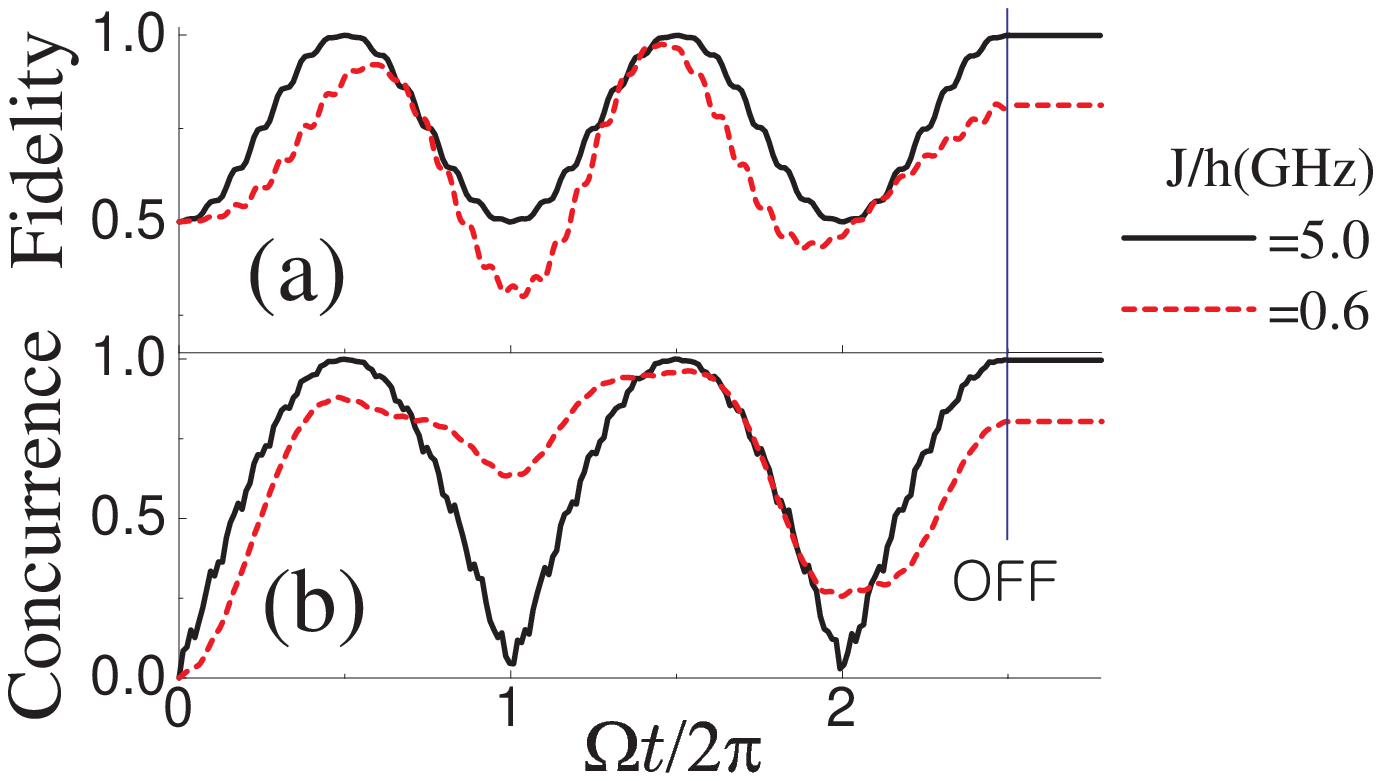}
\vspace{-0.5cm}
%
\caption{ (a) Fidelity and (b) concurrence oscillations with the initial state
$\psi(0)=(|00\rangle+|10\rangle)/\sqrt{2}$ during the CNOT gate
operation for coupled flux qubits. The dashed ($J=0.6
\mathrm{GHz}$) and solid ($J=5 \mathrm{GHz}$) lines
correspond to the weak  and strong  coupling of Fig. \ref{level}, respectively.
At $\Omega t=5\pi$ the coupling is switched off with $f_{a}=f_{b}=0.5$.
}
\label{fidel}
\end{figure}

For a direct comparison of CNOT gate operation performance as the coupling strength varies,
the fidelity and concurrence
\cite{Wootters} oscillations are  plotted as a function of time in
Figs. \ref{fidel}(a) and (b). The fidelity is calculated by the
definition
\begin{eqnarray}
F(t)={\rm Tr}(\rho(t)\rho_{\rm CNOT})/4,
\end{eqnarray}
where $\rho_{\rm CNOT}$ is the matrix for the perfect CNOT operation and
$\rho(t)$ is the truth table amplitude at time $t$ \cite{Plantenberg}.
At $t=0$, $\rho(0)$ is the $4\times 4$ identity matrix and the fidelity has a
finite value  $F(0)=0.5$.
The CNOT operation changes an initial product state into a maximally entangled state.
Thus the maximal entanglement as well as the maximal fidelity
corresponds to the perfect CNOT gate operation. Figure
\ref{fidel} shows that
the deviations of the fidelity and concurrence oscillations diminish as the coupling
strength increases.

At the end of  two-qubit
operations the phase-coupling is switched off with $f_{a}=f_{b}=0.5$.
Since the magnetic dc
pulse for switching-off has a finite rising time, the phase of qubit
state evolves during the time, but this phase evolution is
controllable by manipulating other parameters. The states of qubits
can be detected by shifting the magnetic pulse adiabatically
\cite{Kaku}. At the degeneracy point in Fig. \ref{level}, the averaged current of qubit
states vanishes. Thus, one can apply a finite dc magnetic pulse to
shift the qubits slightly away from the degeneracy point to detect
the qubit current states.

In summary, we study the operation of two coupled flux qubits under a microwave irradiation.
The flux qubits interact  with the threading microwave field by the fluxoid quantization
condition and can be  treated as  magnetic moments for a weak magnetic field
threading the qubit loop.
By using the coupled qubit Hamiltonian
we show that for a strong coupling the microwave-driven CNOT gate can be realized
with a high fidelity.
The strong coupling between flux qubits is obtained by a  phase-coupling scheme.
Introducing the switching SQUIDs
in the phase-coupled qubit results in the renomalization of the magnetic moments of both qubits.
The fluctuation effects of both the static and microwave fields are discussed.

\ack
{The author thanks  Center for
Modern Physics in Chongqing University for hospitality.}

\end{document}